\documentclass[nolinenumbers]{aastex631}

\usepackage{placeins}
\usepackage{subfigure}

\usepackage{longtable}
\usepackage{booktabs}
\usepackage{natbib}
\usepackage{tabularx}
\usepackage{array} 
\usepackage{amsmath} 

\newcommand{\pr}{\textcolor[RGB]{40,150,90}} 

\begin{document}




\title{The Ophiuchus DIsk Survey Employing ALMA (ODISEA): A Unified Evolutionary Sequence of Planet-Driven Substructures Explaining the Diversity of Disk Morphologies}

\author[0000-0002-7625-1768]{Santiago Orcajo}
\affiliation{Facultad de Ciencias Astronomicas y Geofisicas, Universidad Nacional de La Plata, Paseo del Bosque s/n, 1900 La Plata, Argentina}
\affiliation{Instituto de Astrofísica de La Plata (IALP), CCT La Plata-CONICET-UNLP, Paseo del Bosque s/n, La Plata, Argentina}

\author[0000-0002-2828-1153]{Lucas A. Cieza}
\affiliation{Instituto de Estudios Astrofisicos, Facultad de Ingeniería y Ciencias,
Universidad Diego Portales,  Av. Ejercito 441, Santiago, Chile}
\affiliation{Millennium Nucleus on Young Exoplanets and their Moons (YEMS), Chile}

\author[0000-0001-8577-9532]{Octavio Guilera}
\affiliation{Facultad de Ciencias Astronomicas y Geofisicas, Universidad Nacional de La Plata, Paseo del Bosque s/n, 1900 La Plata, Argentina}
\affiliation{Instituto de Astrofísica de La Plata (IALP), CCT La Plata-CONICET-UNLP, Paseo del Bosque s/n, La Plata, Argentina}

\author[0000-0003-2953-755X]{Sebasti\'an P\'erez}
\affiliation{Departamento de Física, Universidad de Santiago de Chile, Av. V\'ictor Jara 3493, Santiago, Chile}
\affiliation{Millennium Nucleus on Young Exoplanets and their Moons (YEMS), Chile}
\affiliation{Center for Interdisciplinary Research in Astrophysics Space Exploration (CIRAS), Universidad de Santiago de Chile, Chile}

\author[0000-0002-3244-1893]{Fernando R. Rannou}
\affiliation{Departamento de Ingenier\'ia Inform\'atica, Universidad de Santiago de Chile, Av. Víctor Jara 3659, Estación Central, Santiago, Chile}
\affiliation{Millennium Nucleus on Young Exoplanets and their Moons (YEMS), Chile}
\affiliation{Center for Interdisciplinary Research in Astrophysics Space Exploration (CIRAS), Universidad de Santiago de Chile, Chile}

\author[0000-0003-4907-189X]{Camilo Gonz\'alez-Ruilova}
\affiliation{Departamento de Física, Universidad de Santiago de Chile, Av. V\'ictor Jara 3493, Santiago, Chile}
\affiliation{Millennium Nucleus on Young Exoplanets and their Moons (YEMS), Chile}
\affiliation{Center for Interdisciplinary Research in Astrophysics Space Exploration (CIRAS), Universidad de Santiago de Chile, Chile}


\author[0009-0007-4878-0252]{Grace Batalla-Falcon}
\affiliation{Instituto de Estudios Astrofisicos, Facultad de Ingeniería y Ciencias,
Universidad Diego Portales,  Av. Ejercito 441, Santiago, Chile}

\author[0000-0002-4314-9070]{Trisha Bhowmik}
\affiliation{Instituto de Estudios Astrofisicos, Facultad de Ingeniería y Ciencias,
Universidad Diego Portales,  Av. Ejercito 441, Santiago, Chile}
\affiliation{Millennium Nucleus on Young Exoplanets and their Moons (YEMS), Chile}

\author[0000-0003-2406-0684]{Prachi Chavan}
\affiliation{Instituto de Estudios Astrofisicos, Facultad de Ingeniería y Ciencias,
Universidad Diego Portales,  Av. Ejercito 441, Santiago, Chile}
\affiliation{Millennium Nucleus on Young Exoplanets and their Moons (YEMS), Chile}

\author[0000-0002-0433-9840]{Simon Casassus}
\affiliation{Departamento de Astronomía, Universidad de Chile, Casilla 36-D, Santiago, Chile }
\affiliation{Data Observatory Foundation, to Data Observatory Foundation, Eliodoro Yañez ˜ 2990, Providencia, Santiago, Chile}
\affiliation{Millennium Nucleus on Young Exoplanets and their Moons (YEMS), Chile}

\author[0009-0009-8115-8910]{Anuroop Dasgupta}
\affiliation{Instituto de Estudios Astrofisicos, Facultad de Ingeniería y Ciencias,
Universidad Diego Portales,  Av. Ejercito 441, Santiago, Chile}
\affiliation{Millennium Nucleus on Young Exoplanets and their Moons (YEMS), Chile}

\author[0000-0002-3244-1893]{Kevin Diaz}
\affiliation{Departamento de Ingenier\'ia Inform\'atica, Universidad de Santiago de Chile, Av. Víctor Jara 3659, Estación Central, Santiago, Chile}
\affiliation{Millennium Nucleus on Young Exoplanets and their Moons (YEMS), Chile}
\affiliation{Center for Interdisciplinary Research in Astrophysics Space Exploration (CIRAS), Universidad de Santiago de Chile, Chile}

\author[0009-0009-4329-2916]{Jos\'e L. Gomez} 
\affiliation{Facultad de Ciencias Astronomicas y Geofisicas, Universidad Nacional de La Plata, Paseo del Bosque s/n, 1900 La Plata, Argentina}
\affiliation{Instituto de Astrofísica de La Plata (IALP), CCT La Plata-CONICET-UNLP, Paseo del Bosque s/n, La Plata, Argentina}

\author[0000-0001-5073-2849]{Antonio S. Hales}
\affiliation{National Radio Astronomy Observatory, 520 Edgemont Road, Charlottesville, VA 22903-2475, USA}
\affiliation{Joint ALMA Observatory, Avenida Alonso de Cordova 3107, Vitacura 7630355, Santiago, Chile}
\affiliation{Millennium Nucleus on Young Exoplanets and their Moons (YEMS), Chile}


\author[0000-0002-1575-680X]{J.M. Miley}
\affiliation{Departamento de Física, Universidad de Santiago de Chile, Av. V\'ictor Jara 3493, Santiago, Chile}
\affiliation{Millennium Nucleus on Young Exoplanets and their Moons (YEMS), Chile}
\affiliation{Center for Interdisciplinary Research in Astrophysics Space Exploration (CIRAS), Universidad de Santiago de Chile, Chile}

\author[0000-0001-8031-1957]{Marcelo M. Miller Bertolami}
\affiliation{Facultad de Ciencias Astronomicas y Geofisicas, Universidad Nacional de La Plata, Paseo del Bosque s/n, 1900 La Plata, Argentina}
\affiliation{Instituto de Astrofísica de La Plata (IALP), CCT La Plata-CONICET-UNLP, Paseo del Bosque s/n, La Plata, Argentina}

\author[0000-0001-8450-3606]{P.H. Nogueira}
\affiliation{Millennium Nucleus on Young Exoplanets and their Moons (YEMS), Chile}

\author[0000-0003-1385-0373]{Mar\'{\i}a Paula Ronco}
\affiliation{Facultad de Ciencias Astronomicas y Geofisicas, Universidad Nacional de La Plata, Paseo del Bosque s/n, 1900 La Plata, Argentina}
\affiliation{Instituto de Astrofísica de La Plata (IALP), CCT La Plata-CONICET-UNLP, Paseo del Bosque s/n, La Plata, Argentina}

\author[0000-0003-3573-8163]{Dary Ruiz-Rodriguez}
\affiliation{National Radio Astronomy Observatory, 520 Edgemont Road, Charlottesville, VA 22903-2475, USA}


\author[0000-0002-5991-8073]{Anibal Sierra}
\affiliation{Mullard Space Science Laboratory, University College London, Holmbury St Mary, Dorking, Surrey RH5 6NT, UK}

\author[0000-0001-9527-2903]{Julia Venturini}
\affiliation{Department of Astronomy, University of Geneva, Chemin Pegasi 51, 1290 Versoix, Switzerland.}

\author[0000-0002-3354-6654]{Philipp Weber}
\affiliation{Departamento de Física, Universidad de Santiago de Chile, Av. V\'ictor Jara 3493, Santiago, Chile}
\affiliation{Millennium Nucleus on Young Exoplanets and their Moons (YEMS), Chile}
\affiliation{Center for Interdisciplinary Research in Astrophysics Space Exploration (CIRAS), Universidad de Santiago de Chile, Chile}

\author[0000-0001-5058-695X]{Jonathan P. Williams}
\affiliation{Institute for Astronomy, University of Hawaii, Honolulu, HI 96822, USA}

\author[0000-0002-5903-8316]{Alice Zurlo}
\affiliation{Instituto de Estudios Astrofisicos, Facultad de Ingeniería y Ciencias,
Universidad Diego Portales,  Av. Ejercito 441, Santiago, Chile}
\affiliation{Millennium Nucleus on Young Exoplanets and their Moons (YEMS), Chile}





\begin{abstract}

Understanding the origin of substructures in protoplanetary disks and their connection to planet formation is currently one of the main challenges in astrophysics.  
While some disks appear smooth, most exhibit diverse substructures such as gaps, rings, or inner cavities, with varying brightness and depth. 
As part of the Ophiuchus Disk Survey Employing ALMA (ODISEA), we previously proposed an evolutionary sequence to unify this diversity, driven by the formation of giant planets through core accretion and subsequent planet-disk interactions.
By combining the disk evolution and planet formation code \textsc{PlanetaLP} with the radiative transfer code {\sc radmc-3D}, we have now reproduced the key aspects of the proposed evolutionary sequence.  
Starting with a smooth disk (like e.g., WLY 2-63), we modeled the evolution of a fiducial disk with a 1 Jupiter-mass planet at 57 au. Within a few hundreds of orbits, a narrow gap forms, resembling ISO-Oph~17. By $\sim$0.1 Myr, the gap widens, and dust accumulates at the cavity edge, producing a structure similar to Elias 2-24. At $\sim$0.4 Myr, the disk evolves further into a morphology akin to DoAr 44, characterized by a smaller inner disk and a brighter inner rim. By $\sim$1 Myr, the system transitions to a single narrow ring, resembling RXJ1633.9–2442.
This line of work strongly supports the planetary origin of substructures and enables the possibility of identifying a population of planets that is currently beyond the reach of more direct detection techniques. 

\end{abstract}


\section{Introduction} \label{sec:intro}

High-resolution images of massive ($\gtrsim$ 10 M$_{\rm Jup}$) protoplanetary disks have shown that they contain a plethora of substructures \citep{2015ApJ...808L...3A,2018ApJ...869L..41A,2018ApJ...869...17L}. The most common type of substructures are gaps, rings, and cavities \citep{2020ARA&A..58..483A}. Although other types of substructures, such as spiral arms \citep{2016Sci...353.1519P} and nonaxisymmetric dust traps \citep{2013Sci...340.1199V,Marino2015ApJ...813...76M}, can also be found, they are much less frequent. 
Cavities, gaps, and rings seem to be shallower and less common in young embedded systems \citep{2023ApJ...954..110O}. This might be due to a lower intrinsic incidence of substructures at very young ages or to optical depth effects (or to a combination of both). 

Even before ALMA, planet-formation has been one of the leading explanations for disk substructures \citep[see][for a pre-ALMA review]{2011ARA&A..49...67W}. 
However, the fact that some rings and gaps seem to appear at very early ages ($\lesssim$ 1 Myr) at up to  $\gtrsim$ 100 au from the central star, as in the case of HL~Tau \citep{2015ApJ...808L...3A},  presents significant challenges 
for the two models of planet formation which are most favored, core accretion and gravitational instability \citep{Pollack+1996, Boss1997}. 
In particular, core accretion is highly inefficient at large distances from the star \citep[e.g.][]{2020A&A...638A...1M}, and most disks with gaps are not massive enough to be gravitationally unstable or show signs of gravitational instability, like the formation of spiral arms \citep{2014MNRAS.444.1919D}.  
As a result, additional alternative origins are often considered for these substructures, including snow-lines \citep{2015ApJ...806L...7Z}, magneto-hydrodynamic effects \citep{2015A&A...574A..68F}, and other instabilities such as the secular gravitational instability \citep[e.g.][]{2011ApJ...731...99Y,2014ApJ...794...55T} and the viscous ring-instability \citep{2018A&A...609A..50D}. 
Nevertheless, the direct detection of two protoplanets within the gap of the PDS 70 disk
\citep{2018A&A...617A..44K,2019NatAs...3..749H} demonstrates that planets can indeed form at tens of au within a few Myr and gives credence to a direct link between substructures and planets.

In recent years, simulations of planet-disk interactions have successfully reproduced several of the substructures observed in protoplanetary disks with remarkable accuracy. Notable examples include the triple-ring system in the transition disk HD169142, which reveals the presence of a single mini-Neptune in the outer disk \citep{2019AJ....158...15P}, the two mini-Saturns shepherding a crescent-shaped dust trap in HD163296 \citep{2023ApJ...945L..37G}, and the two giant planets carving the rings observed in V4046~Sgr \citep{2022MNRAS.510.1612W}. These cases provide compelling evidence that planet-disk interactions play a fundamental role in shaping the intricate architectures of protoplanetary disks.

In the planet-formation framework, and as part of the ODISEA (Ophiuchus DIsk Survey Employing ALMA) project \citep{2019MNRAS.482..698C,2019ApJ...875L...9W,2020MNRAS.496.5089Z},  we have proposed an evolutionary sequence in which the diversity of substructures observed in disks could be understood in terms of giant planet formation through core accretion and simple dust evolution, without invoking additional physics \citep[][C21 hereafter]{2021MNRAS.501.2934C}. 
\defcitealias{2021MNRAS.501.2934C}{C21}

The Ophiuchus molecular cloud is particularly useful for studying disk evolution because it contains a significant population of embedded sources (mostly in the L1688 cluster) with ages $\lesssim$ 1~Myr
\citep{2009ApJS..181..321E} and a more distributed population of young stellar objects with ages up to $\sim$ 6 Myr \citep{2020AJ....159..282E}. 

The sequence proposed by \citetalias{2021MNRAS.501.2934C} was based on 3-5 au resolution images at 1.3 mm of the 15 brightest disks in Ophiuchus \citep[10 of them observed as part of ODISEA and 5 as part of the DSHARP Large Program,][]{2018ApJ...869L..41A}.  
The samples included 3 embedded sources and 12 Class II sources with very diverse IR Spectral Energy Distributions, consistent with ``full",  ``pre-transition", and ``transition" disks.\footnote{See C21 for detailed definitions of SED classes.} 
The proposed sequence had 5 distinct stages (see fig. 10 in \citetalias{2021MNRAS.501.2934C}):  \\

\textbf{Stage I)} Embedded disks with shallow or no obvious substructures, corresponding to an epoch in which protoplanets are not massive enough to carve noticeable gaps in the disks.   \\

\textbf{Stage II)}  Disks with relatively narrow, but clear gaps and rings, indicating the growth of protoplanets. \\ 

\textbf{Stage III)} A rapid widening of the gaps due to the sudden growth in the mass of some planets when they acquire their gaseous envelopes. This stage includes the rapid accumulation of dust at the outer edges of the gaps (the inner rims of the outer disks) due to the strong pressure bumps \citep{2012A&A...545A..81P} caused by the
giant planets that recently formed. \\  

\textbf{Stage IV)}  Dust filtration at the edges of the cavities resulting in dust-depleted inner disks. The millimeter dust from the outer disks efficiently drifts in \citep{2007A&A...469.1169B} and accumulates at the edges of the gaps.  \\

\textbf{Stage V)}  Eventually,  the dusty inner disks drain completely onto the stars, and the outer disks become narrow rings (or collections of narrow rings).   \\ 

Here, we combine the disk evolution and planet formation code \textsc{PlanetaLP} \citep{Guilera+2017,2017MNRAS.471.2753R, Guilera+2019, Guilera+20, 2020A&A...644A.174V, Guilera+2021, Ronco+2024} with the radiative transfer code {\sc radmc-3D} \citep{2012ascl.soft02015D} 
in order to reproduce the proposed sequence.
Our fiducial disk is similar to Elias 2-24 in terms of size and mass \citep{2017ApJ...851L..23C}. We introduce a planet in the disk and follow the time evolution of the gas and dust components. 
We then create synthetic images of the different epochs, which can be compared to real ALMA images.  
To coarsely explore the parameter space,
this evolution was repeated varying some model parameters (e.g., planet mass, disk mass, and disk viscosity).  

In Section~\ref{sec:model},  we describe the experimental setup, including the properties adopted for our fiducial disk and the application of the \textsc{PlanetaLP} code to compute the dust evolution under the influence of planet-disk interactions.  
In Section~\ref{sec:results}, we explore how the different parameters of the model affect the observed substructures and evolve our fiducial model over several millions of years.  
In Section~\ref{discussion}, we discuss our numerical results in the context of ALMA high-resolution surveys and their implications for planet formation theory.  
We also consider the limitations of our current models and discuss future improvements. 
A summary of our main conclusions is provided in Section~\ref{summary}.   

\section{Models and numerical setup}
\label{sec:model}

\subsection{The fiducial disk}

As mentioned above, our fiducial disk is based on Elias 2-24, a system that is at the middle of the sequence proposed by \citetalias{2021MNRAS.501.2934C} (corresponding to Stage III, in which the envelope of a giant planet has recently been accreted).  
The initial gas surface density of the disk is parameterized as:  

\begin{equation}
    \Sigma_{g} = \Sigma_{g}^{0}\left(\frac{R}{R_{c}}\right)^{-\gamma} e^{-\left(\frac{R}{R_{c}}\right)^{2-\gamma}},
    \label{Eq1}
\end{equation}

\noindent
where we use the characteristic radius ${R_{c}}$ = 120 au and the power-laws index $\gamma$ =  0.8,
as estimated by \citet{2017ApJ...851L..23C} from radiative transfer modeling of low (0.2$^{\prime\prime}$) 
resolution ALMA data.
The $\Sigma_{g}^{0}$ value was adjusted to reach the initial disk mass of 0.1 M$_{\odot}$ estimated by \citet{2017ApJ...851L..23C}. 
Following \citet{2018ApJ...869L..47Z}, which is based on high-resolution (0.04$^{\prime\prime}$) imaging of the system, we place a 1 M$_{\rm Jup}$ mass planet at 57 au from the star, for which we adopt a mass of 0.8 M$_{\odot}$. 
We use the Gaia distance of $139\pm1$~pc \citep{2023A&A...674A...1G}.
We also ran an additional model with a Mars-mass (0.0003 M$_{\rm Jup}$) planet to represent Stage I (embedded disks with shallow or no obvious substructures). 
Our fiducial disk has an $\alpha$ viscosity of $10^{-3}$, and an initial dust-to-gas ratio of 0.02 (see Table~\ref{Tab:ParamList}). 
%
We adopt this initial dust-to-gas mass ratio that is a factor of 2 higher than the standard 0.01 value because this allows us to match the integrated flux of Elias 2-24 at 1.3 mm 
($\sim$0.3 Jy). 
The other alternative would be to further increase the initial total disk mass, which is already very high.

\begin{table}[h]
    
    \caption{
   The parameters and initial conditions of our fiducial model,  shown in \textcolor{blue}{\textbf{blue}}, and  the additional values explored for such quantities. 
    }
    \centering        
            \begin{tabular}{llllll}
            \toprule
            Parameter  & \multicolumn{5}{c}{Value} \\ 
            \midrule
            $M_{p} [M_{\rm Jup}]$ & 0.3 & 0.5 & \textcolor{blue}{\textbf{1.0}} & 3.0 & 5.0 \\ 
            $\alpha$ & $1.0\times 10^{-4}$ &  $5.0\times 10^{-4}$ & \textcolor{blue}{\textbf{1.0$\times$10$^{-3}$}} & $2.5 \times 10^{-3}$ &  $5 \times 10^{-3}$ \\
            dust-to-gas ratio  & 0.005  & 0.01 & \textcolor{blue}{\textbf{0.02}} & 0.05 &  \nodata  \\
            $M_{\rm disk} / M_{\rm star}$ &  0.01  & 0.05 & \textcolor{blue}{\textbf{0.1}} & 0.2 &   \nodata   \\  
            \bottomrule
        \end{tabular}
\label{Tab:ParamList}
\end{table}

\subsection{Dust growth and evolution under planet-disk interactions}

We apply the global planet formation model \textsc{PlanetaLP} to study the dust growth and evolution under the influence that a giant planet generates in the gaseous disk. Regarding the protoplanetary disk --the emphasis of this study--, our model simulates the evolution of a 1D
axisymmetric gaseous disk influenced by viscous accretion and the planet perturbations, which in turn influences the evolution of the dust disc.
The evolution of the dust and the pebbles is modeled as outlined by \citet{Guilera+20}, implementing a discrete size distribution with 200 bins spanning from 1~$\mu m$ to a specified maximum size. Within each radial bin, dust particles grow to a maximum size constrained by processes such as coagulation, radial drift, and fragmentation \citep{Birnstiel+12}. We employ mass-weighted mean drift velocities and mass-weighted mean Stokes numbers to determine the time evolution of the solid surface density by solving an advection-diffusion equation. We also consider that the dust/pebbles properties change at the water ice-line assuming that silicate particles have a threshold fragmentation velocity of 1 m/s inside the water-ice line, while the ice-rich particles beyond the water-ice line have a fragmentation velocity of 10 m/s \citep{Gundlach2015}.

A key aspect of this Letter is the inclusion of the effect of a gap generated by an already-formed giant planet on the evolution of the dust in the disk. We modified {\sc PlanetaLP} to account for dust growth and transport in the presence of this gap. Specifically, we incorporated how the pressure bump that forms at the outer edge of the gap affects the dust dynamics. This addition builds on the work of \citet{2020ApJ...889...16D} by allowing us to investigate how a giant planet influences the retention, accumulation, and radial drift of dust particles. The pressure structure imposed by the gap alters the dust distribution, potentially leading to particle trapping and the formation of ring-like features --a common morphology observed by ALMA in dust continuum--, a process not included in the previous implementations of {\sc PlanetaLP}. Further details of the model implementations are provided in Appendix~\ref{PLP}.

\subsection{Radiative Transfer Modeling and Synthetic Images} 

{\sc radmc-3D} \citep{2012ascl.soft02015D}\footnote{\url{https://www.ita.uni-heidelberg.de/~dullemond/software/radmc-3d/index.php}} is one of the most widely used radiative transfer codes for studying protoplanetary disks. We use this code to convert the surface density profiles generated by {\sc PlanetaLP} into synthetic images at 1.3 mm, matching the wavelength of the high-resolution ODISEA and DSHARP data. 
We have adapted the output \pr{of} \textsc{PlanetaLP} to be compatible with the input requirements of {\sc radmc-3D} (see App.~\ref{RT}).
This allows us to perform radiative transfer modeling and ray tracing of the \textsc{PlanetaLP} models, which can then be compared to the ALMA images. For further details of the radiative transfer calculations, see Appendix~\ref{RT}. We also refer the reader to previous works (e.g., \citealt{2019AJ....158...15P} and \citealt{2019MNRAS.486..304B}) that employ similar procedures to generate ALMA synthetic observations from hydrodynamic simulations of dust and gas using Lagrangian particles with \textsc{radmc-3D}.

To perform realistic comparisons, we created visibility sets of {\sc radmc-3D}  images with the {\tt ft} task within the {\sc CASA} (Common Astronomy Software Applications) package \citep{2007ASPC..376..127M,2022PASP..134k4501C}. 
These visibility sets have the same $u-v$ coverage as the real observations of Elias 2-24.  
Using the {\tt setnoise} task of the {\tt Simulator} tool within CASA, we corrupted the synthetic visibilities to approximate the thermal noise of the observations. 
We then reconstructed the images from the visibilities in the standard fashion (with the {\tt tclean}  algorithm, also  within {\sc CASA}). 
We used multi-scale {\tt tclean}, with briggs weighting, a robust value of 0.5, and 1000 iterations,
resulting in a beam size of 
$\sim$0.04$^{\prime\prime}$.
We note that not all ODISEA and DSHARP objects were observed in exactly the same way; therefore, some differences related to $u-v$ coverage and noise levels are expected between some images and models.
%

\section{Results} \label{sec:results}

\begin{figure}[h]
    \centering
    \includegraphics[width=1.\columnwidth]{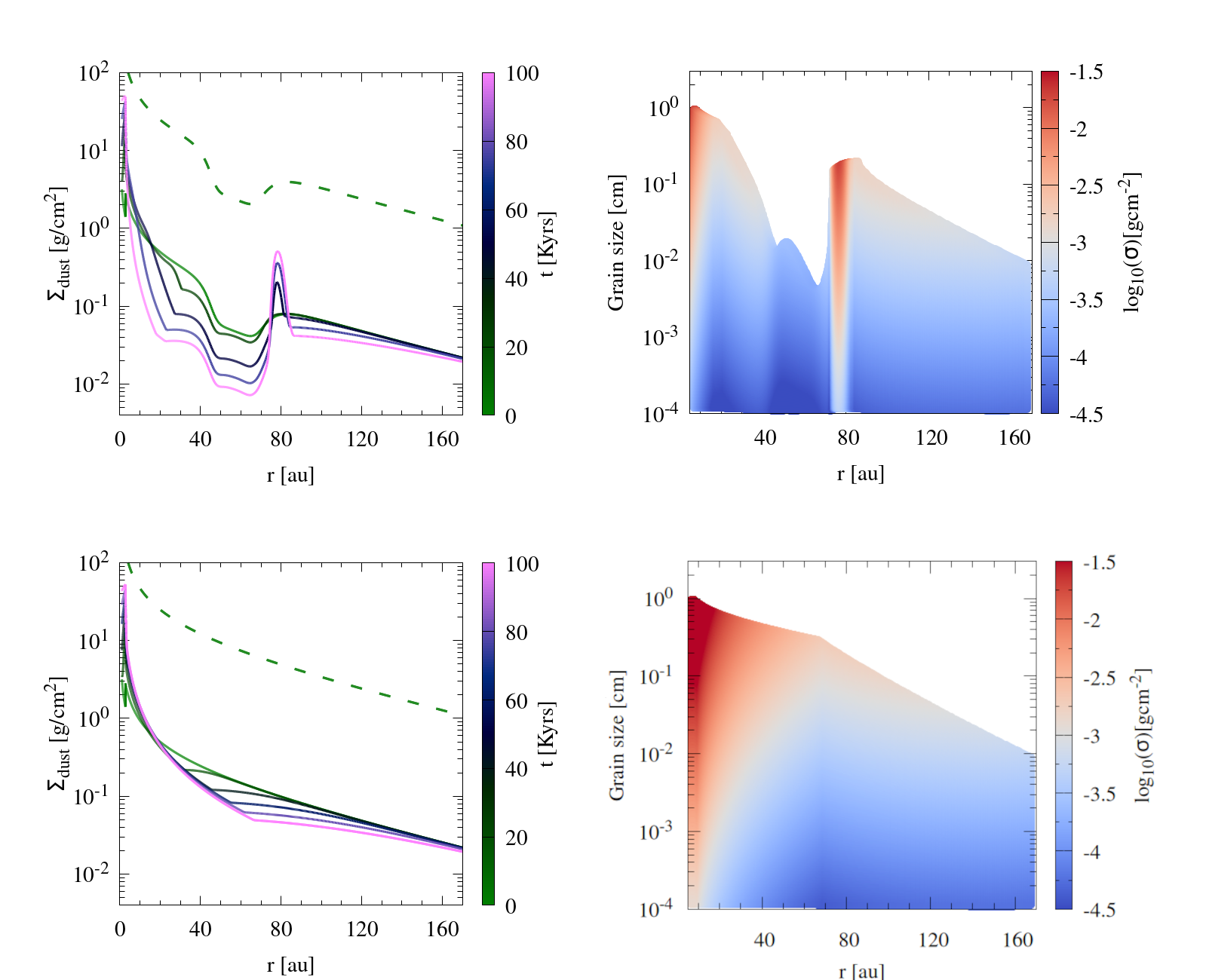}
   \caption{
   \textbf{Top panels:}
   models with 1 M$_{\rm Jup}$ planet.
   The left panel shows the surface density profiles of our fiducial model. The dashed green line corresponds to the initial gas surface density, while the solid lines indicate the evolution of the dust surface density in steps of 25000 years after a 1~M$_{\rm Jup}$ planet has been injected at 57 au. The right panel shows the maximum grain size as a function of radius of the fiducial model after 0.1 Myr evolution. The color scale corresponds to the integrated surface density at the given radius. As expected, the dust accumulates and grows at the outer edge of the gap.
   \textbf{Bottom panels:} same as top panels, but for the model with a Mars-mass planet, which we use to represent Stage I.  While no gap is produced, an inflection point in the surface density profile and the grain-size distribution is seen close to the  location of the planet.}
    \label{DustGrain}
\end{figure}

\subsection{Fiducial model} 

In this Letter, we do not model the growth of the planet core and the accretion of the envelope,  we simply inject the fully grown giant planet in our fiducial disk and let the system evolve.  
In Figure \ref{DustGrain}, we show the first 0.1~Myr of evolution of our fiducial disk after inserting the 1~M$_{\rm Jup}$ planet.
Observing the evolution of the dust disk, we find that as time passes, the gap deepens, the inner disk becomes narrower, and dust accumulates at the outer edge of the gap at around 80~au (see left panel), which matches the main characteristics of Elias 2-24.  
We also find that, due to grain growth and size-dependent radial drift, dust grains at the edge of the gap are significantly larger ($\gtrsim$ 1 mm) than in the rest of the outer disk (see right panel). 
However, we find that the dust does not grow to the levels seen in the inner disk (size $\gtrsim$ 1 cm). This is consistent with multi-frequency analyses of the Elias 2-24 disk (Chavan et al. in prep). 
We note here that in the inner regions of the disk, the maximum grain size is constrained by fragmentation, whereas in regions farther out, it is governed by radial drift (see \ref{app_dust}). The transition between these two regimes appears as a change in the slope of the dust surface density radial profiles, approximately between 20 and 35 au \citep[this effect is studied e.g. by][]{2016A&A...594A.105D, Rosotti+2019}.
We also show the evolution of the model with the Mars-size planet (bottom panels). We find that, as expected, no gap is formed.  However, a break (inflection point) is seen in both the surface density profile and the grain size distribution close to the location of the planet. This is interesting because inflection points are common in the brightness profiles of protoplanetary disks when observed at high-resolution in millimeter continuum (\citetalias{2021MNRAS.501.2934C}) and tend to be present in very young embedded objects, even when they do not show deep gaps with clear local minima (Bhowmik et al.
in prep.).
Our modeling results suggest that such inflection points might be caused by low-mass protoplanets.

\subsection{Exploration of the Parameter Space} 

To investigate the effect of each model parameter on the resulting surface density profile, we vary one parameter at a time, over the ranges of values listed in Table~\ref{Tab:ParamList}, and compare the results to our fiducial model. 
For this, we evolve each system for 0.1 Myr, corresponding to $\sim$230 orbits of the giant planet assumed to be in a circular and coplanar orbit.  

\subsubsection{The effect of planet mass and viscosity} 

\begin{figure}[h]
    \centering
\includegraphics[width=1.0\columnwidth]{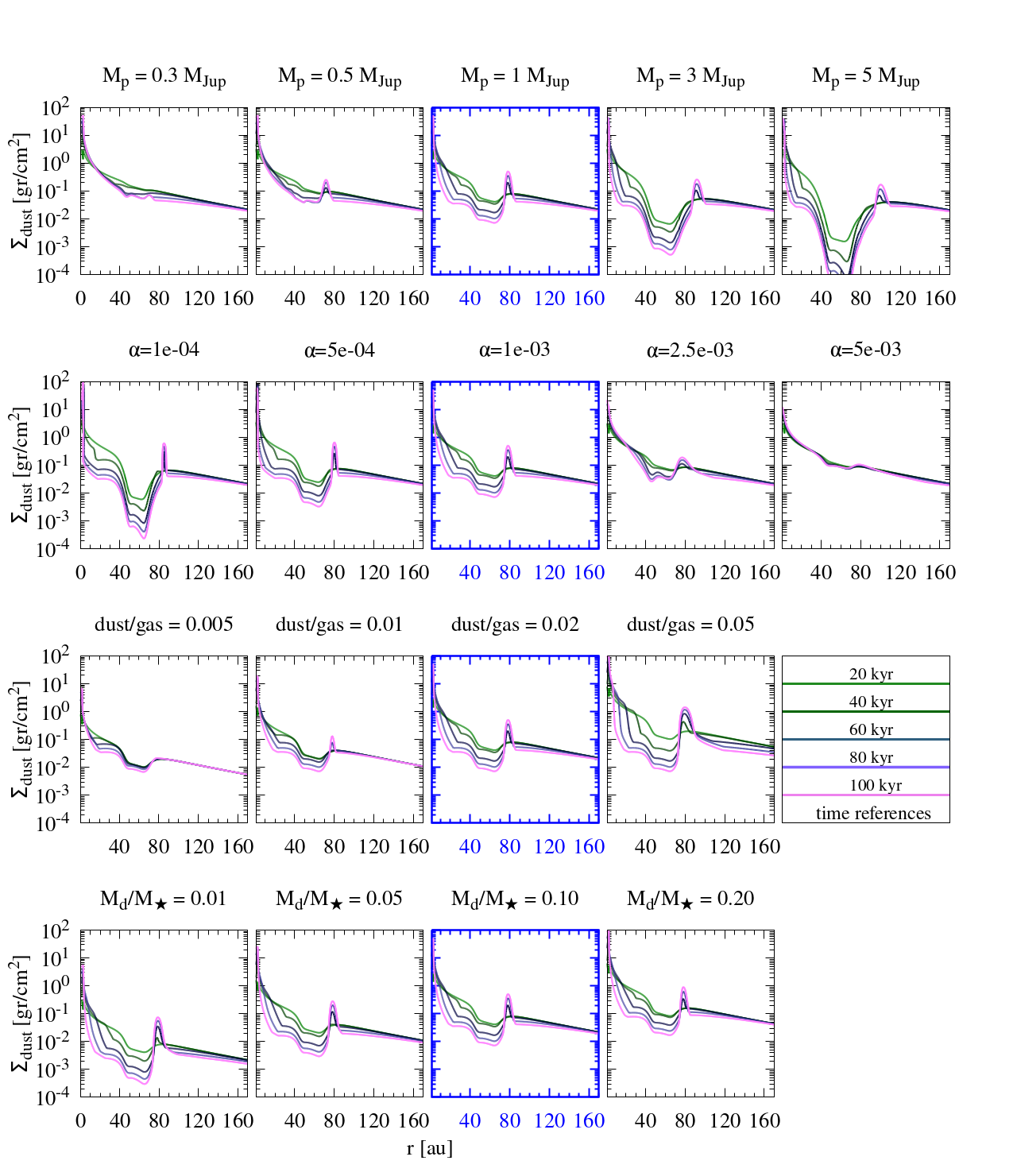}
   \caption{From the top to bottom, variation of $M_{p}$, $\alpha$, dust-to-gas mass ratio, and M$_{disk}$/M$_{star}$, compared with fiducial model \textcolor{blue}{(blue mark)}.}
    \label{Mp-a-eff}
\end{figure}

In the top panel of Fig.~\ref{Mp-a-eff},  we show the effect of the planetary mass $M_{p}$ on the dust surface density, leaving all other parameters as in the fiducial model. 
As expected, we verify that, the larger the planetary mass, the deeper and wider the gap.
We further find that larger planets produce somewhat
wider rings where the dust accumulates in the outer disk, although the effect is much less evident.    
In addition, the peak location of the ring is observed a larger distances as $M_{p}$ increases. 
In the upper-middle panel of Fig.~\ref{Mp-a-eff} we also verify that, the higher the $\alpha$ value, the more difficult it is to form the gap and the bright ring. Both features become much less discernible when $\alpha$ =  5 $\times$ $10^{-3}$. The inverse effect occurs for small alpha ($\alpha$ = $10^{-4}$).
We note that planet mass and disk viscosity are somewhat degenerate, in the sense that different combinations of parameters can produce gaps with similar widths and depths.  

\subsubsection{The effect of disks mass and dust to gas mass ratio}

We find that as the dust-to-gas ratio increases, the main effect is the widening of the ring and an increase in the dust accumulation at the edge of the gap, but without significantly affecting the radial position of the peak (Fig. \ref{Mp-a-eff}, lower-middle panel). 
As expected, we observed a general increase in the surface density profile across the disk as we increased the dust-to-gas ratio.
In Fig. \ref{Mp-a-eff}, (bottom panel), we observe that, as expected, as the mass of the disk increases, the surface density also increases.
As is the case with the dust-to-gas ratio, the position of the ring is not affected. 
As both the total disk mass and the dust-to-gas mass ratio increase the total mass of solids in the disk, these parameters are partially degenerate. 
However, the effects related to grain growth and radial drift are not identical for both parameters,  and the differences are more noticeable at the edge of the gap.  

\subsection{Morphological Temporal Evolution}  

\begin{figure}[h]
    \centering
    \includegraphics[width=1.\columnwidth]{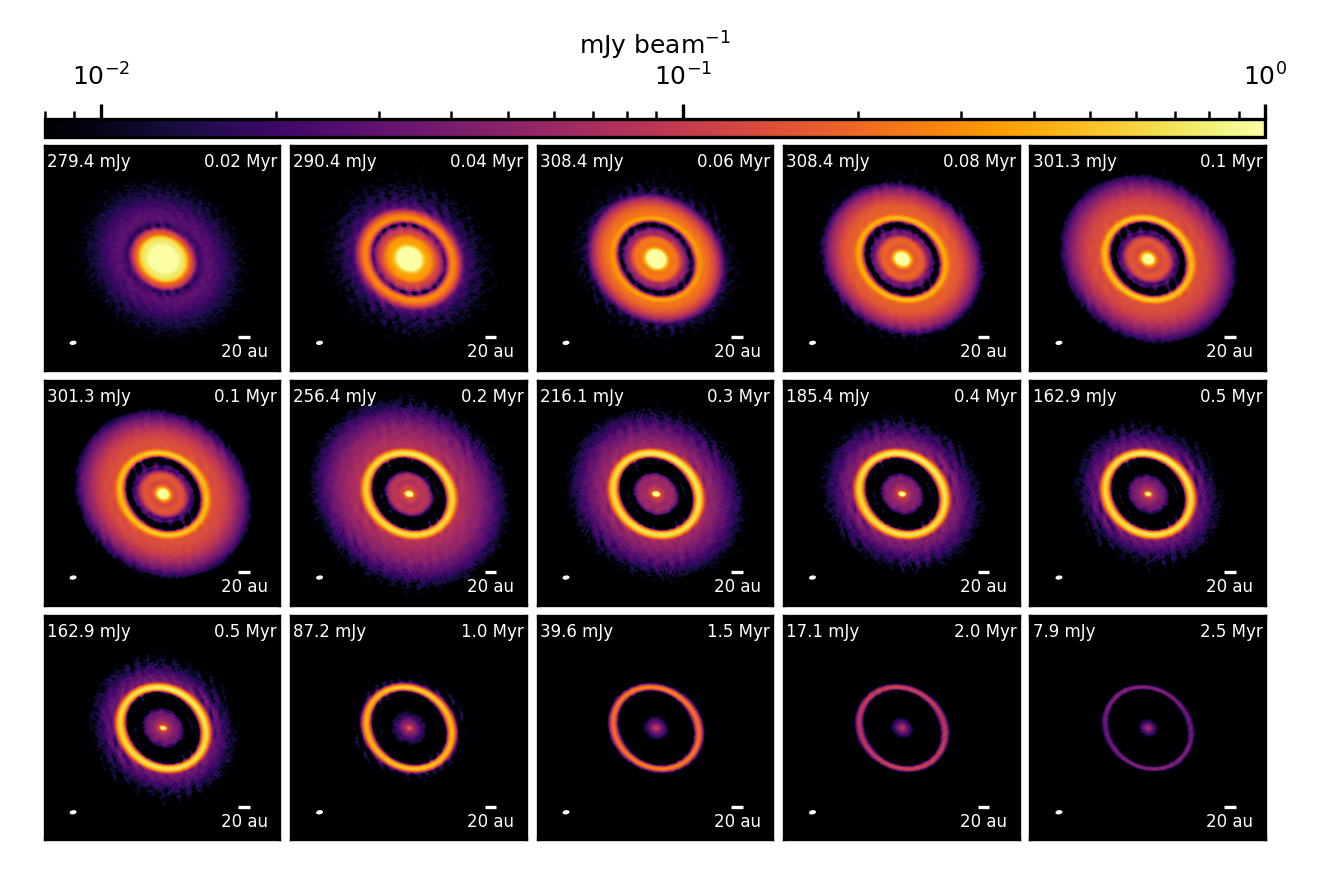}
   \caption{The temporal evolution for fiducial model with the 1~M$_{\rm Jup}$ planet, from 20~kyrs to 2.5~Myrs. 
    The top five panels show steps of 20~kyrs,
    the middle panels  correspond to steps of   100~kyrs,
    and the lower panels illustrate steps of 500~kyrs.
    Each panel corresponds to the synthetic image at 1.3 mm of the  \textsc{PlanetaLP} model, produced via radiative transfer and ray-tracing with \textsc{RADMC-3D}. 
   }
    \label{Evolution}
\end{figure}

The application of {\sc radmc-3D} to \textsc{PlanetaLP} outputs allows to explore the morphological evolution of our fiducial model, as it would be seen by ALMA.   
In Figure \ref{Evolution}, we show synthetic images of the fiducial model between 20 kyr and 3 Myr.
The panels in the first row have steps of only 20 kyr to focus on the rapid development of the gap.
The panels in the middle row correspond  to steps of 100 kyr to illustrate the evolution of the inner and the outer disks. 
Finally,  the panels in the bottom row, show the longer-term evolution of the system with steps of 500 kyr.
This figure clearly demonstrates that  planet-driven evolution can dramatically alter the morphology of a 
single system. 
In the following section, we discuss our numerical results in the context of the diversity of structures observed in molecular clouds and the evolutionary sequence proposed by \citetalias{2021MNRAS.501.2934C}.

\section{discussion}
\label{discussion}
\subsection{Towards an evolutionary sequence of substructures in
massive disks}

\begin{figure}[h]
    \centering
    \begin{minipage}[b]{1.\textwidth}
        \centering
     \includegraphics[width=1.0\columnwidth]{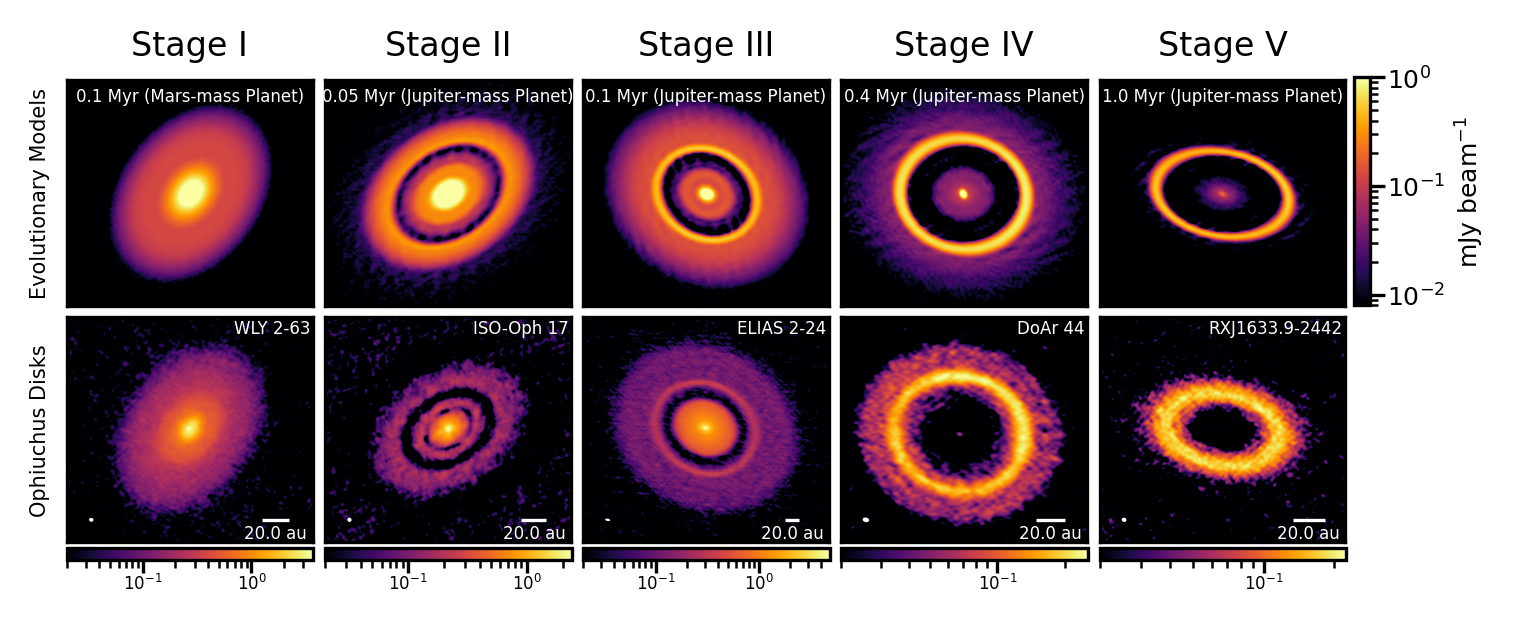}       
        \caption{
        The 
        \textbf{top panels} 
        show simulated images of our models. The first panel corresponds to the model with the Mars-mass planet at 0.1 Myr.  The rest of the panels correspond to 0.05, 0.1, 0.4, and 1 Myr after the 1~M$_{\rm Jup}$ planet has been injected.
        The sizes, inclinations, and position angles have been adjusted to match those of the real ALMA images of  WLY 2-63, ISO-Oph 17, Elias 2-24, DoAr 44, and RXJ1633.9–2442, shown in the \textbf{bottom panels}. 
        The images of the models and the real ALMA images have similar u-v coverages and imaging parameters. 
        The similarities in morphology are striking, considering the simplicity of our models.
        The very short timescales of Stages II is due to the fact that we do not model the slow growth of the core of the giant planet. 
        A 42 second animation is available in the online journal. The video shows models of the evolution of the fiducial disk over a period of 1 Myr, combined with real ALMA images illustrating each stage.
        }
        \label{fig:evolution}
    \end{minipage}
\end{figure}

Planetary systems show a remarkable diversity of architectures, ranging from compact systems with multiple rocky planets ($\lesssim$ 1 M$_{\oplus}$) within $<$ 0.1 au, like TRAPPIST I \citep{2017Natur.542..456G}, to very extended systems with multiple massive planets
($\sim$5-10 M$_{\rm Jup}$) at tens of au as in the case of HR 8799 \citep{2008Sci...322.1348M,2010Natur.468.1080M}.  
This diversity, at least in part, is likely to be due to the range of disk properties observed in molecular clouds. 
In this context, it is important to keep in mind that the evolutionary sequence proposed by \citetalias{2021MNRAS.501.2934C} is based on the top 5$\%$ of the brightest disks present in Ophiuchus. 
Such disks have several things in common: they are large, massive, and capable of forming giant planets. 
Therefore, it is not entirely surprising if they evolve in a similar way. 
Also, Ophiuchus has the particularity of having young stellar objects with a wide range of ages, including several embedded Class I and Flat Spectrum sources \citep{2009ApJS..181..321E}, and a large number of Class II sources with diversified structures, consistent with different stages of evolution.  
These characteristics render Ophiuchus a particularly valuable laboratory to study disk evolution and planet formation.  
\citetalias{2021MNRAS.501.2934C} used WLY 2-63\footnote{WLY 2-63, also known as IRS 63, is not completely smooth. It shows very shallow gaps consistent with early stages of planet formation \citep{2020Natur.586..228S}.}
, ISO-Oph 17,  Elias 2-24, DoAr 44, and RXJ1633.9–2442 (among other disks) to illustrate Stages I, II, III, IV, and V  of the proposed evolutionary sequence.  
In Figure \ref{fig:evolution}, we compare 
%
our models to the observations of these objects. We use the model with a Mars-mass planet at 0.1 Myr to reproduce Stage I and the models with the  1~M$_{\rm Jup}$ planet at 0.05, 0.1, 0.4, and 1 Myr to reproduce the rest of the Stages.
We find that our simple models are able to reproduce the main features of each Stage: 
the lack of clear gaps at Stage I, the formation of narrow gaps at Stage II,   the wide gap and the accumulation of dust at the outer edge of the gap at Stage III,  the brightening of the inner rim, and the dimming of the inner disk in Stage IV, and the formation of a single narrow ring at Stage V.  
The fact that a simple model can reproduce five different morphologies is remarkable, especially considering that no fine-tuning of the model has been performed to match each disk except for the mass of the planet in Stage I.   
In reality, each system most likely has a unique planetary architecture with multiple planets in addition to the one giant planet in our fiducial model that drives the morphology of the disk. 
The presence of multiple planets is expected in very massive disks capable of forming gas giants (e.g., the Solar System has several planets in addition to Jupiter).
Differences in the number, orbits, and masses of the additional planets in each system are likely to explain the small discrepancies observed in Figure \ref{fig:evolution} (e.g., the size of the inner disk, or the presence of additional narrow gaps). At Stage V a significant difference between our fiducial model and RXJ1633.9–2442 is the persistence of an inner disk, which does not dissipate completely. 
We speculate that this could also be due to the presence of inner planets in  RXJ1633.9–2442 and/or the effect of photoevaporation. 
Photoevaporation should also result in the final dispersal of the primordial disk and the transition to the debris disk phase 
\citep{2014prpl.conf..475A}. 
However, investigating the effect of additional planets and photoevaporation is beyond the scope of this Letter. Another possibility to explain the lack of an inner disk in the  image of  RXJ1633.9–2442 is related to dust fragmentation efficiency, which determines the amount of grains large enough to emit at ALMA wavelengths
\citep{2018ApJ...864L..26B}.

\subsection{Age considerations}
 
Estimating the absolute ages of young stellar objects is notoriously difficult and is usually done by comparing the stellar temperature and luminosities to different evolutionary tracks, which can easily result in age differences of more than 1~Myr \citep{2014prpl.conf..219S}.  This approach is particularly problematic for deeply embedded objects for which stellar luminosities are highly uncertain.  However, even with these limitations we can still reasonably assess whether the ages of the real systems shown in Fig. \ref{fig:evolution} are consistent with the planet formation stages we propose. WLY-2~63 is an embedded source with a Flat Spectrum SED and an estimated age of $<$ 0.5~Myr based on its bolometric luminosity of just 290 K \citep{2020Natur.586..228S}. ISO-Oph~17 and Elias~2-24 are two Class II objects with “full” SEDs located in the high-extinction  regions of the L1688 cluster (Av $>$ 8 mag; C21),  with estimated ages of 1-2 Myrs \citep{2020AJ....159..282E,2018ApJ...869L..41A}. DoAr 44 is a Class II source located in the L1699 region of the Ophiuchus cloud. It has a pre-transition SED, low extinction (Av $\lesssim$ 1 mag;   
 \citetalias{2021MNRAS.501.2934C}), and moderate accretion \citep[6.5 $\times$ $10^{-9}$ M$_{\odot}$/yr;][]{2020A&A...643A..99B}. Therefore,  it is reasonable to conclude that DoAr 44 is slightly older than ISO-Oph 17 and  Elias 2-24. At the end of the sequence is RXJ1633.9–2442, which has a transition disk SED, a very low accretion rate of
 $\lesssim$ 10$^{-10}$ M$_{\odot}$/yr  and an estimated age  of $\gtrsim$ 2-5 Myr \citep{2012ApJ...752...75C}. Given all of the above, we conclude that the evolutionary sequence is consistent with the relative ages of the sources, even though we also note that a perfect correlation between stellar age and structure Stage is not expected. The timing of planet formation is likely to vary from system to system and the duration of the planet structure Stages should also depend on the location and mass of the planet.

\subsection{Caveats and future prospects}\label{caveats}

\begin{figure}[h]
    \centering
    \begin{minipage}[b]{1.0\textwidth}
        \centering
     \includegraphics[width=1.0\columnwidth]{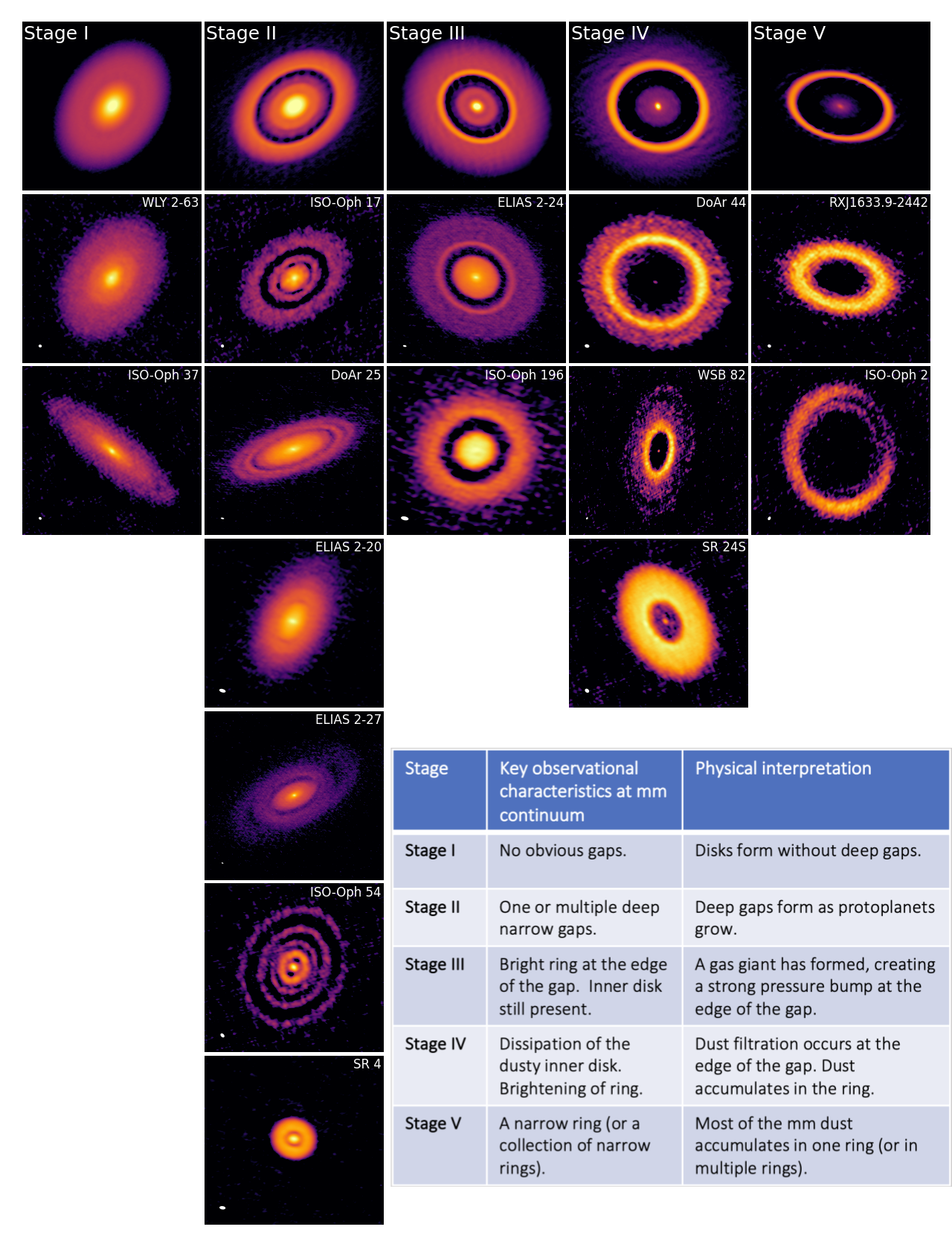}       
        \caption{The top panels show our models, the same as in Fig.\ref{fig:evolution}, illustrating each on of the five stages. 
        The rest of the panels contain the 15 brightest disks in Ophiuchus studied by \citetalias{2021MNRAS.501.2934C} in the Stage (column) in which they have  been classified.  
        The figure illustrates the key characteristic features of each Stage and also the observed diversity (see \ref{caveats} for discussion).
        %
        %
        }
        \label{fig:full-sample}
    \end{minipage}
\end{figure}

The goal of this Letter is to test the overall sequence of planet-driven evolution of substructures of massive disks proposed by \citetalias{2021MNRAS.501.2934C}, rather than to find precise matches between specific systems and our models.
For this purpose, we employed the \textsc{PlanetaLP} code, originally developed for other applications, with minimal modifications---primarily the key addition of a gap induced by a recently formed giant planet and adjustments to ensure compatibility with the input requirements of {\sc radmc-3D}. 
We have only performed a coarse exploration of the parameter space, but have already obtained very promising results.  
Figure \ref{fig:evolution} clearly shows that an evolving protoplanetary system can look notably different at different epochs as the result of planet-disk interactions and dust evolution.  
As expected,  the morphology of the dusty disk is a function of planetary architecture \emph{and} time.   
Even though some rare disk structures might be unrelated to planets \citep[e.g., complex structures might be caused by stellar flybys;][]{2020MNRAS.491..504C},  planet-driven evolution  might account for most of the morphological diversity observed in protoplanetary disks.
%

We do not aim to explain all disks with gaps and rings with our simple model, especially not those systems where the substructures might be due to multiple giant planets.  
In Fig. \ref{fig:full-sample} we include the full sample of Ophiuchus disks studied by \citetalias{2021MNRAS.501.2934C}. 
The figure shows that objects in Stage II   might have multiple narrow gaps, while our model only has one.  
Similarly,  not all objects in Stage IV look the same. This Stage is characterized by the dissipation of the inner disk and the further accumulation of dust at the edge of the gap.   
Our fiducial model at Stage IV looks very similar to DoAr 44,
but different from WSB 82 and SR 24S.  
We hypothesize that this is also due to the fact that our model has a single planet. In the absence of additional planets in the outer disk, the dust drifts unimpeded, resulting in a narrow outer disk very quickly (only 0.2 Myr in our fiducial model).   
Additional outer planets are likely needed to explain the structures of WSB 82 and SR 24S.

 \textsc{PlanetaLP} and {\sc radmc-3D} are a very powerful tools for this type of work, which can be expanded to include multiple planets.  
\textsc{PlanetaLP}  also has the capabilities of including  planet-growth and migration, two processes that are expected to be important for the planet-driven evolution of protoplanetary disk, but have not been considered in this Letter.  
For instance, in our fiducial model, the formation of the gap (Stage II) happens extremely fast (within 0.05 Myr). 
This very short timescale is most likely an artifact of injecting a fully grown giant planet. 
In practice,  the core of the planet is expected to grow over a longer period of time \citep{Pollack+1996, Guilera+2010, Morbidelli+2015}, resulting in a much slower evolution of the gap.
The fact that many disks have narrow gaps that lack the bright edges seen around Elias 2-24 and in more evolved disks (see Fig. \ref{fig:full-sample})
suggests that Stage II is one of the longest phases in the sequence and that Stage III is one of the shortest.  
To better match the demographic statistics, we plan to implement a more realistic modeling of the first two Stages in the future,
by including the slow formation of the planetary cores.  
Also, \textsc{PlanetaLP} provides surface density profiles for dust populations of different grain sizes, a feature that will allow us to explore the morphology of a disk as a function of wavelength (via {\sc radmc-3D}) and investigate whether the low occurrence of deep gaps at early stages is due to optical depth effects or to a lower intrinsic incidence \citep{2023ApJ...954..110O}.

 Models of disk evolution and planet-disk interactions have multiple free parameters to describe the disk (mass, size, dust-to-gas mass ratio, and viscosity) and the planetary architecture (number of planets, semi-major axes, and masses). With such a large number of parameters, the exploration of the parameter space becomes difficult and time-consuming. Therefore, exact matches between the models and the observations cannot be expected, unless the parameter space can be explored in a very efficient way. Also, several degeneracies between parameters are known to exist (e.g., different combinations of planet mass and disk viscosity can produce a similar gap). 
 One significant advantage of 1-D models of protoplanetary disk evolution like \textsc{PlanetaLP} is computational efficiency.
 These models allow for a detailed treatment of dust growth, fragmentation, and radial transport, while significantly reducing computational costs compared to full hydrodynamic simulations. 
 %
 3-D hydrodynamical codes such as \textsc{Fargo3D} \citep{2000A&AS..141..165M, BLM2016} capture more detailed physical interactions, including instabilities and non-axisymmetric structures ---that could allow, for instance to include warped inner disks in our model such as DoAr44's \citep{Arce2023MNRAS.526.2077A}. 
 Hydrodynamical codes could also allow one to investigate the detailed effects of planet migrations, which could also contribute to the diversity of substructures \citep{2019ApJ...884..178W,2019AJ....158...15P} as shown by other works without performing radiative transfer calculations \citep[e.g.][]{2019MNRAS.482.3678M}.   
 However,  1-D models remain an invaluable tool for exploring large parameter spaces, disk demographics, and understanding the fundamental mechanisms driving dust evolution.  Still, even with 1-D models,  obtaining precise parameters and associated uncertainties requires advanced techniques for the exploration of the parameter space, like Markov chain Monte Carlo (MCMC) simulations. The implementation of such an exploration algorithm is also beyond the scope of this Letter, but is highly desirable for future advances in this line of research.

\subsection{Implications for planet formation and demographic context}

Assuming a direct connection between dust substructures and planets opens up the possibility of using high-resolution ALMA images to constrain the properties of the underlying population of planets in protoplanetary disks \citep[e.g.,][]{2018ApJ...869L..47Z}.  
The results presented herein provide very strong support for such an assumption. 
In this context, the most straightforward implication of our work is that planet formation might be extremely efficient in some systems, even at large radii.  
For instance,  ISO-Oph 54 is an embedded (Class I) disk with four gaps at 12, 32, 48, and 72 au from the star (see Fig. \ref{fig:full-sample}). Elias 2-24 itself is also believed to be very young (age $\lesssim$ 1 Myr, \citealt{2018ApJ...869L..41A}).  
Explaining the short timescale for forming planetary cores suggested by the observations is indeed one of the biggest challenges for planet-formation models in the ALMA era
\citep{2023ASPC..534..717D}.

The evolutionary sequence proposed by \citetalias{2021MNRAS.501.2934C} is mostly qualitative,  but with additional modeling,  quantitative criteria can be developed to define each Stage based on images or radial profiles. 
Such criteria can then be systematically applied to large samples of high-resolution data to observationally constrain their relative duration (Bhowmik et al. in prep.). The relative numbers of objects at each  Stage shown in Fig. \ref{fig:full-sample} can already provide some hints because it represents a flux-limited sample.
As mentioned above, Stage II has the most objects and probably has the longest duration. Stage III has only two members, Elias 2-24 and ISO-Oph~196, and is likely to be very short.  This Stage is characterized by a wide gap, the accumulation of dust at the edge of the gap,  and the presence of an inner disk, which is mostly dissipated by stage IV.   
The short duration of Stage III is not too surprising considering that it corresponds to an epoch very soon  ($\sim$0.1 Myr) after the gas giant is fully formed.
The gas giant isolates the inner disk from the outer disk.  
The mm-sized grains in the inner disk quickly drift in, but the inner disk is no longer fed by mm grains from the outer disk. This is so because the pressure bump produced by the giant planet results in efficient dust filtration at the edge of the gap and only grains smaller than $\sim$10-100 micron can cross the gap \citep{2006MNRAS.372L...9R,2012ApJ...755....6Z}.
%

It is interesting to note that our proposed evolutionary sequence is also in agreement with the broad demographic trends obtained by ALMA Large Programs that observed samples of different mean ages.
The eDisk (Early Planet Formation in Embedded Disks) survey targeted very young embedded sources with ages  $<$1 Myr and detected mostly structures consistent with our Stage I \citep{2023ApJ...954..110O}. The DSHARP (Disk Substractures at High Angular Resolution Project) focused on young  Class II sources with a mean age of $\sim$1 Myr and mainly detected disks resembling our Stages II and III \citep{2018ApJ...869L..41A}. Finally, the AGE-PRO (ALMA Survey of Gas Evolution in Protoplanetary Disks) program recently found a high incidence of disks with large central dust cavities, similar to our Stage V, in Upper Sco targets, which have ages in the 2-6 Myr range (Vioque et al. 2025, in press).

When discussing disk demographics, it is also important to recall that the proposed evolutionary sequence is based on the brightest, most massive disks in Ophiuchus. In particular,  all disks shown in Fig. \ref{fig:full-sample} have estimated dust masses higher than $\sim$40 M$_{\oplus}$ and are tens of au in radius (\citetalias{2021MNRAS.501.2934C}).
Our numerical results thus mainly apply to massive disks capable of forming giant planets. 
However, most disks in Ophiuchus and other star-forming regions might lack the mass needed to form gas giants.
In fact, most ODISEA targets have estimated dust masses lower than $\sim$3 M$_{\oplus}$ \citep{2019ApJ...875L...9W} and are smaller than $\sim$15 au in radius \citep{2025arXiv250115789D}.  
According to exoplanet statistics
\citep{2016MNRAS.457.2877G}, there are more rocky planets than stars in the Galaxy, which implies that even faint and compact disks should form rocky planets and might have planet-induced gaps that are only detectable if observed at  $\lesssim$ 1~au resolution, as seen in TW~Hydra \citep{2016ApJ...820L..40A,2019MNRAS.484L.130M}.  
Such a high resolution pushes the current ALMA limits but should be within reach of the ngVLA \citep{2024ApJ...965..110W}, which will also have the advantage of observing at longer wavelengths and probing lower optical depths.

\section{Summary and Conclusions}
\label{summary}
We combined detailed simulations of dust and gas evolution using \textsc{PlanetaLP} with radiative transfer calculations and simulations of millimeter-wavelength observations to reproduce the key features of the evolutionary sequence proposed by \citetalias{2021MNRAS.501.2934C}. In our framework, the prominent substructures observed in massive protoplanetary disks arise primarily from giant planet formation via core accretion. The gravitational influence of these nascent planets redistributes the gas density, which in turn drives the subsequent evolution of the dust. This integrated approach not only validates the \citetalias{2021MNRAS.501.2934C} scenario but highlights the crucial interplay between planet formation and disk structure evolution in shaping young planetary systems.  In particular,  it provides strong support for the following aspects of the proposed sequence:  
\\

1) A single system, at different times, might show the characteristic morphology of each of the evolutionary stages in the sequence. \\

2) Planets can carve deep, yet narrow gaps, as commonly seen in many Class II sources. \\

3) The morphology of the Elias 2-24 disk corresponds to an epoch very shortly ($\lesssim$ 0.1 Myr) after the moment when the planet gained its gaseous envelope and became a giant. This epoch is characterized by the formation of a wide gap and the rapid accumulation of millimeter-sized dust on the outer edge of the gap (the inner rim of the outer disk) due to the strong pressure bump produced by the recently formed planet. \\

4) The sequence progresses with the dissipation of the inner disk (within the orbit of the giant planet) and the further accumulation of solids at the edge of the gap, due to rapid inward drift of the dust.  \\

5) In the final stage of the sequence, all dust from the outer disk accumulates in a narrow ring.
\\

6) Over time, bright rings gradually fade, potentially marking the transition to a debris disk phase.  \\

The combination \textsc{PlanetaLP} with {\sc radmc-3D} has allowed us to perform direct comparisons between  high-resolution ALMA images and models of disk evolution and planet-disk interactions. 
Our models with a single planet in a massive (0.1 M$_{\odot}$) disk matches well the morphology of several disks in Ophiuchus (WLY 2-63, ISO-Oph 17, Elias 2-24, DoAr 44, and RXJ1633.9), spanning all five Stages proposed by \citetalias{2021MNRAS.501.2934C}; however, reproducing the full diversity of substructures requires a much wider exploration of the parameter in terms of disk properties and planetary architectures.       
Nonetheless, our results strongly suggest that the most common substructures seen in the dust continuum images of protoplanetary disks (gaps and rings) can be understood in terms of the formation of planets  \textit{and} that the subsequent
evolution of these substructures is driven by planet-disk interactions and dust evolution.  
Using  \textsc{PlanetaLP}, our initial results could be expanded to include multiple planets and the formation of the initial structures produced by sub-giant proto-planets (activating planet-growth and even migration). 
However, the results presented herein already have important implications for the field because if the proposed sequence is correct, it would mean that Jupiter-mass planets can form at many tens of au from a star within $\lesssim$ 1 Myr. 
Also, we emphasize that this line of work enables the possibility of identifying a population of planets that is currently below other more direct detection techniques.
The number of parameters in simulations such as these is large, as they include both disk and (multi)planet properties; nevertheless, in principle, a more efficient exploration of the parameter space (e.g., via MCMC) can be implemented to constrain the underlying planetary architecture of any system where high-quality high-resolution ALMA data are available.      
%

\section{acknowledgments}
We thank to constructive comments from the anonymous referee, which have helped to significantly improve the Letter.
S.O., O.M.G., M.P.R., M.M.B. and J.L.G. are partially supported by PIP-2971 from CONICET (Argentina) and by PICT 2020-03316 from Agencia I+D+i (Argentina).
L.A.C. acknowledges support from ANID, FONDECYT Regular grant number 1241056, Chile.
L.A.C., S.P., F.R., K.D., C.G-R., P.W., and A.Z. acknowledge support from the Millennium Nucleus on Young Exoplanets and their Moons (YEMS), ANID - NCN2021\_080 and NCN2024\_001, Chile. 
S.P., F.R., and K.D. acknowledge support from FONDECYT Regular grant 1231663. 
P.W. acknowledges support from FONDECYT grant 3220399.
M.P.R. is partially supported by PICT-2021-I-INVI-00161 from ANPCyT, Argentina. S.O., O.M.G., M.P.R., M.M.B. and J.L.G. also thank Juan Ignacio Rodriguez from IALP for the computation managing resources of the Grupo de Astrof\'{\i}sica Planetaria de La Plata. 
J.V. acknowledges support from the Swiss National Science Foundation (SNSF) under grant PZ00P2\_208945.
This paper makes use of the following ALMA data: ADS/JAO.ALMA \# 2016.1.00484.L and \# 2018.1.00028.S. 
ALMA is a partnership of ESO (representing its member states), NSF (USA),
and NINS (Japan), together with NRC (Canada), NSC and ASIAA (Taiwan), and KASI (Republic of Korea), in cooperation with the Republic of Chile. The Joint ALMA Observatory is operated by ESO, AUI/NRAO, and NAOJ. The
National Radio Astronomy Observatory is a facility of the National Science Foundation operated under cooperative agreement by Associated Universities, Inc. 
%

\vspace{5mm}
\facilities{ALMA}

\software{CASA, PlanetaLP, RADMC-3D}

\appendix

\section{Overview of PLANETALP as applied in this study}
\label{PLP}
\subsection{Gas component}

The gaseous disk evolves in time by viscous accretion. The gas surface density of the disk $\Sigma_{\rm g}$ obeys the classical 1D radial diffusion equation \citep{Pringle1981}: 
\begin{eqnarray} 
  \frac{\partial \Sigma_{\rm g}} {\partial t}= \frac{3}{r}\frac{\partial}{\partial r} \left[ r^{1/2} \frac{\partial}{\partial r} \left( \nu \Sigma_{\rm g} r^{1/2}  \right) \right], 
\label{eq1-app-planetalp}
\end{eqnarray}
$t$ and $r$ being the temporal and radial coordinates, and $\nu= \alpha c_s H_{\rm{g}}$ the turbulent viscosity, given by the dimensionless parameter $\alpha$ \citep{SS73}. The local sound speed $c_s$ is given by 
\begin{equation}    c_{s}=\sqrt{\frac{\gamma k_{B}T}{\mu \, m_{H}}},
\label{eq2-app-planetalp}
\end{equation}
where $\gamma$ = 7/5 is the adiabatic constant, $k_{B}$ is the Boltzmann constant, $\mu$ = 2.3 is the mean molecular weight of molecular gas assuming a typical H-He disk composition, and $m_{H}$ is the mass of a proton. Since we aim to study the evolution of disk structures far from the central star, we assume a vertically isothermal disk, where the mid-plane temperature follows the profile of a passive disk \citep{Ida2016}. The scale height of the disk is set to $\rm H_{\rm g}=c_{s}/\Omega_{k}$, where $\Omega_{k}$ is the Keplerian frequency. 
We solve Eq.~\ref{eq1-app-planetalp} implicitly in finite differences using a radial grid with 2000 bins between 0.1~au and 1000~au logarithmically equally spaced, using zero torques as boundary conditions.

Finally, the volumetric gas density and the pressure at the disk midplane are given by,
\begin{equation}
    \rho_{\rm g}=\frac{\Sigma_{\rm g}}{\sqrt{2\pi} \rm H_{\rm g}},
\end{equation}
\begin{equation}
    \rm P_{\rm g}=c_{s}^{2}\rho_{\rm g}.
\end{equation}

\subsubsection{The gap induced by a planet}

We incorporated the generation and effect of a gap opened by a giant planet in \textsc{PlanetaLP} to study the dust growth and evolution in such 
 gaseous disk. The implementation is based on the work of \cite{2020ApJ...889...16D}, according to which the gas density profile including the gap induced by the planet is given by:
\begin{equation}
    \Sigma_{\rm g} = \frac{\Sigma_{\rm unper}}{1+\frac{0.45}{3\pi}\frac{q(r)^{2}\mathcal{M}^{5}}{\alpha}\delta(q(r))}
    \label{eq-planet-gap}
\end{equation}
where $\Sigma_{\rm unper}$ is the unperturbed gas surface density, i.e. without the presence of the planet, and 

\begin{eqnarray}
    q(r) &=&\frac{q}{\left[1+D^{3}\left(\left(\frac{R}{r_{p}}\right)^{1/6}-1\right)\right]^{1/3}}, \\
    \mathcal{M} &=& \left(\frac{r_{p}}{H_{g}}\right)^{5}, \\
    D &=& 7\frac{\mathcal{M}^{3/2}}{\alpha^{1/4}}, \\
    q &=& \frac{M_\mathrm{P}}{M_\star}.
\end{eqnarray}

The novelty here is that 
\begin{equation}
    \delta(q) =
    \begin{cases} 
        1, & q < q_{\mathrm{NL}} \\ 
        (q / q_{\mathrm{NL}})^{-1/2} + (q / q_w)^3, & q > q_{\mathrm{NL}}
    \end{cases}
\end{equation}
where $q_{NL} = 1.04\mathcal{M}^{-3}$, and $q_{w} = 34 q_{NL}\sqrt{\alpha\mathcal{M}}$. \cite{2020ApJ...889...16D} explain that $\delta(q)$ counteracts the effect that occurs when the mass ratio starts to become important because for $q>10^{-4}$ the planetary torques start to deviate from linearity. The breaking moment would occur for $q_{NL}$. On the other hand, $q_{w}$ is the mass ratio at which the gap density reaches $\sim$2\% of the gas density.

To solve the time evolution of the gas surface density via Eq.~\ref{eq1-app-planetalp} we use always $\Sigma_{\rm unper}$. Then, at each time step we compute $\Sigma_{\rm g}$ using Eq.~\ref{eq-planet-gap} to mimic the induced gap by the giant planet on the disk. Finally, the dust growth and evolution is solved using the perturbed gas disk as background. We note that the introduction of an already formed planet and the instantaneous appearance of a gap in the gaseous disk do not generate any numerical problem to solve the advection-diffusion equation that models the dust evolution (Eq.~\ref{eqAdvDif}, see next section)

\subsubsection{Dust component}
\label{app_dust}

As in \citet{Guilera+20}, we compute the dust growth and evolution considering a discrete size distribution using 200 size bins, from $1~\mu m$ to a maximum size at each radial bin. 
Such maximum size is limited by coagulation, drift, and fragmentation \citep{Birnstiel+12}. The dust change of properties at the water-ice line --defined as the place where the disk temperature at the midplane equals 170 K--. We adopt that inside the water-ice line, silicate particles have a fragmentation velocity of 1~m/s, and that ice-rich dust particles beyond the water-ice line have a fragmentation velocity of 10~m/s \citep{Gundlach2015}.

Thus, the maximum particle size at a given time is given by:
\begin{equation}
  r_{\text{d}}^{\text{max}}(t)= \min(r_{\text{d}}^0 \, \exp(t/\tau_{\text{growth}}), \, r_{\text{drift}}^{\text{max}}, \, r_{\text{frag}}^{\text{max}},
  r_{\text{ddf}}^{\text{max}})
  \label{eq1-app112-planetalp}
\end{equation}
where $r_{\text{d}}^0= 1~\mu\text{m}$ is the initial dust size, $\tau_{\text{growth}}$ is the collisional growth timescale given by 
\begin{equation}
  \tau_{\text{growth}}= \frac{1}{Z\Omega_{\text{k}}}
  \label{eq2-sec2-1-0}
\end{equation}
being $Z= \Sigma_{\rm d}/\Sigma_{\rm g}$ the dust-to-gas ratio. The maximum size of dust particles limited by radial drift is given by 
\begin{equation}
  r_{\text{drift}}^{\text{max}}= f_{\text{d}} \frac{2\Sigma_{\text{d}}v_{\text{k}}^2}{\pi \rho_{\text{d}} c_s^2} \left| \frac{d\,\ln P_{\text{g}}}{d\,\ln r} \right|^{-1}, 
  \label{eq3-sec2-1-0}
\end{equation}
where $f_{\text{d}}= 0.55$ \citep{Birnstiel+12}, $v_{\text{k}}$ is the Keplerian velocity, and $\rho_{\text{d}}$ is the mean dust density, 
taking values of $\rho_{\text{d}}$ =$3~\text{g}/\text{cm}^3$ and $\rho_{\text{d}}$=$1~\text{g}/\text{cm}^3$ inside and outside the ice line, respectively. The maximum size of dust particles limited by fragmentation is given by: 
\begin{equation}\label{rmax}
  r_{\text{frag}}^{\text{max}}= f_{\text{f}} \frac{2 \Sigma_{\rm g} v_{\text{th}}^2}{3\pi \rho_{\text{d}} \alpha c_s^2}, 
\end{equation}
where $f_{\text{f}}= 0.37$ \citep{Birnstiel11}, and  $v_{\text{th}}$ is the fragmentation threshold velocity. Finally, if the viscosity in the disk becomes very low, fragmentation occurs through differential drift, given by \citep{Birnstiel+12}:
\begin{equation}
  r_{\text{ddf}}^{\text{max}}= \frac{4 \Sigma_{\rm g} v_{\text{th}}v_{\text{k}}}{c_s^2 \pi \rho_{\text{d}}} \left| \frac{d\,\ln P_{\text{g}}}{d\,\ln r} \right|^{-1}. 
  \label{eq5-sec2-1-0}
\end{equation}

The time evolution of the dust surface density, $\Sigma_{\rm d}$ obeys an advection-diffusion equation, 
\begin{equation}
  \frac{\partial}{\partial t} \left(\Sigma_{\text{d}}\right) + \frac{1}{r} \frac{\partial}{\partial r} \left( r \,v_{\text{drift}} \,\Sigma_{\text{d}} \right) - \frac{1}{r} \frac{\partial}{\partial r} \left[ r \,D \,\Sigma_{\rm g}  \frac{\partial}{\partial r} \left( \frac{\Sigma_{\text{d}}}{\Sigma_{\rm g}} \right) \right] = \dot{\Sigma}_{\text{d}},  
  \label{eqAdvDif}
\end{equation}
where $\dot{\Sigma}_{\text{d}}$ represents in this study the sink term due pebble sublimation when they cross the water-ice line. $D= \nu / ( 1 + \text{St}^2 )$ is the dust diffusivity \citep{YoudinLithwick2007}, and $\text{St}= \pi \rho_{\text{d}} {r}_{\text{d}} / 2 \Sigma_{\rm g}$ the mass weighted mean Stokes number of the dust size distribution. $r_{\text{d}}$ represents the mass weighted mean radius of the dust size distribution, given by
\begin{equation}
  r_{\text{d}}= \frac{ \sum_i \epsilon_i r_{\text{d}}^i }{ \sum_i \epsilon_i },  
  \label{eq2-sec2-1}
\end{equation}
being $r_{\text{d}}^i$ the radius of the dust particle of the species $i$, and $\epsilon_i= \rho^{i}_{\text{d}}/\rho_{\text{g}}$ the ratio between the volumetric dust density of the species i and the volumetric gas density \citep[see][]{Guilera+20}. We note that because the mean size is mass-weighted, the maximum and mean sizes are very similar, as shown in \citet[][see App. A]{Guilera+20}, . To compute the weighted mean drift velocities of the pebbles population ($v_{\text{drift}}$), we follow the same approach as in \citet{Guilera+20} considering the reduction in the pebble drift velocities by the dust-to-gas back-reaction due to the increment in the dust-to-gas ratio. Again, Eq.~\ref{eqAdvDif} is solved with a fully implicit finite differences method in a grid of 2000 radial bins between 0.1~au and 1000~au logarithmically equally spaced using zero dust density as boundary conditions. 

Although we do not account for the vertical evolution of the dust particle density distribution, we determine the total volume density for each dust size based on the scale height of each size particle at a given radial distance $h_d(r)_i$, given by \citep{YoudinLithwick2007}
\begin{equation}
h_d(r)_i= \rm H_{\rm g} \left(\frac{\alpha}{\alpha + {\rm St(r)}_i}\right)^{1/2},
\end{equation}
where ${\rm St(r)}_i$ is the Stoke number of the size particle $i$ at the distance $r$. Thus, the volumetric density distribution of the dust in cylindrical coordinates for each particle size is given by  \citep{Pinilla+2021}

\begin{equation}
    \rho_d(r, \theta, z)_i = \frac{\Sigma_{\rm d}(r)_i}{\sqrt{2\pi} h_d(r)_i} 
    \exp \left( - \frac{z^2}{2 h_d(r)_i^2} \right),
    \label{eq-vol-density-dust}
\end{equation}

where $\Sigma_{\rm d}(r)_i$ represents the dust surface density for each particle size at the distance $r$.

\subsection{Radiative transfer}
\label{RT}

To bridge the output of \textsc{PlanetaLP} with the input requirements of \textsc{radmc-3D}, we first consolidate the volumetric density distribution of the 200 dust size bins provided by \textsc{PlanetaLP} via Eq.~\ref{eq-vol-density-dust} into 9 logarithmically spaced size bins between $1~\rm \mu m$ and the $r_{\rm d}^{\rm max}$ at each $r,\theta,z$ bin. This binning scheme maintains an approximately equal number of particles per decade in size, significantly reducing the computational cost of the radiative transfer calculations while preserving the features of the dust distribution.
The resulting bins in the \textsc{radmc-3D} input have the following minimum and maximum sizes in microns: (1.0, 5.0), (5.0 10.0), (10.0, 50.0), (50.0, 100), (100, 500), (500, 1000), (1000, 5000) (5000, 10000), (10000, 100000). 
To ensure that the vertical structure of the disk is adequately resolved, we choose the number of colatitude cells based on the disk’s vertical extension. Since \textsc{radmc-3D} operates in cartesian or spherical coordinates rather than cylindrical, we resample our volumetric dust density fields onto a Cartesian grid.
Dust opacities for each size bin are computed independenyly using the {\tt optool} application \citep{2021ascl.soft04010D} and a dust grain composition consisting of 99\% astro silicates and 1\% graphite.
With these opacities and the adapted dust density fields, we perform both Monte Carlo thermal calculations and ray-tracing simulations. In our Monte Carlo runs, we utilize $10^8$ photon packages and adopt the stellar spectrum of Elias~2-24. The synthetic images are generated under the assumption that the disk is located at the distance of Elias~2-24---a reasonable approximation given that all the disks in our sample lie within the Ophiuchus cloud.
This radiative transfer procedure enables us to directly compare our \textsc{PlanetaLP} models with ALMA observations, thereby providing a robust framework for testing our scenarios of disk evolution and planet formation.


\bibliography{main}{}
\bibliographystyle{aasjournal}

\end{document}